\documentclass[10pt,article]{amsart}
\usepackage{amssymb,url}

\usepackage{amssymb}
\usepackage{amsthm}
\usepackage{amsmath}

\usepackage{geometry}
\newcommand\tr{\textup{Tr}\,}

\newcommand{\C}{\mathbb{C}}

\newcommand{\N}{\mathbb{N}}

\newcommand{\R}{\mathbb{R}}

\newcommand{\irH}{\mathcal{H}}
\newcommand{\irK}{\mathcal{K}}

\newcommand{\irP}{\mathcal{P}}

\begin{document}

\title{An elementary proof for the non-bijective version of Wigner's theorem}
\author{Gy.~P.~Geh\'er}
\address{Bolyai Institute\\
University of Szeged\\
H-6720 Szeged, Aradi v\'ertan\'uk tere 1, Hungary}
\address{MTA-DE "Lend\"ulet" Functional Analysis Research Group, Institute of Mathematics\\
University of Debrecen\\
H-4010 Debrecen, P.O. Box 12, Hungary}
\email{gehergy@math.u-szeged.hu}
\urladdr{\url{http://www.math.u-szeged.hu/~gehergy/}}

\begin{abstract}
The non-bijective version of Wigner's theorem states that a map which is defined on the set of self-adjoint, rank-one projections (or pure states) of a complex Hilbert space and which preserves the transition probability between any two elements, is induced by a linear or antilinear isometry. We present a completely new, elementary and very short proof of this famous theorem which is very important in quantum mechanics. We do not assume bijectivity of the mapping or separability of the underlying space like in many other proofs.
\end{abstract}

\maketitle

\begin{center}
AMS (2010): 46C05, 46C50.

PACS (2010): 11.30.-j, 03.65.Ta.

\end{center}


\section{Introduction}

E.~P.~Wigner was the first who introduced the definition of the so called \emph{symmetries} in quantum mechanics. A symmetry is a transformation of a quantum structure that preserves a certain quantity or relation. Specifically, a symmetry on the set of rank-one projections that preserves the transition probability between any two elements is usually called a \emph{Wigner symmetry}. It was stated in \cite{Wi} in 1931 that every bijective Wigner symmetry is induced by a unitary or an antiunitary operator. This result is known as Wigner's theorem. The more general, non-bijective version is also well-known and it states that every (not necessarily bijective) Wigner symmetry is induced by a linear or an antilinear isometry.

It is worth mentioning that U.~Uhlhorn proved a certain generalization of the bijective Wigner theorem in \cite{U}. Namely he only assumed that the bijective transformation preserves orthogonality in both directions. However, in that proof the bijectivity assumption and the condition that the underlying space is at least three-dimensional are crucial. We note that the general version of Wigner's theorem cannot be extended in such way, in fact, there are very simple examples which show that usually a transformation that preserves orthogonality in both directions is not induced by a linear or an antilinear isometry.

If we consider a unitary or an antiunitary operator on the underlying Hilbert space, then it obviously defines a Wigner symmetry. The hard part is to prove the reverse direction. A very simple example for a unitary transformation is a \emph{rotation} of the coordinate system. The well-known \emph{time reversal operator} cannot be linear, but it can be realized as an antiunitary operator (see \cite{Wi}). On a separable Hilbert space the \emph{shift operator} 
\[ {\bf S}(x_1,x_2,\dots) = (0,x_1,x_2,\dots) \]
provides a linear isometry which is not unitary (since it is not bijective) and it corresponds to a (non-bijective) Wigner symmetry.

In quantum mechanics, a complex Hilbert space is associated to every quantum system where usually separability is assumed. However, there are certain situation in quantum field theory where it is necessary to consider non-separable Hilbert spaces as well. A very simple example is the following: if we consider an infinite tensor product of at least two-dimensional Hilbert spaces, then it will be non-separable (see e.~g.~\cite[page 86--87]{SW}). Therefore it is relevant to consider Wigner symmetries on non-separable spaces as well.

Concerning the history of the theorem, Wigner himself did not give a rigorous mathematical proof. The first such proof (for the bijective case) was given by J.~S.~Lomont and P.~Mendelson, thirty-two years later in \cite{LM}. One year after that V.~Bargmann gave another proof in \cite{Ba}. Several other proofs were given so far for the bijective and non-bijective versions, see for instance \cite{CVLL,Ch,Gy,Molnar1,Molnar2,Ra,SA}.

In \emph{Physics Letters A} recently three elementary and short proofs were given: two in \cite{SMCS} and one in \cite{Mo}. However, in the first paper bijectivity of the mapping and separability of the underlying Hilbert space are assumed, and in the second letter two times differentiability is assumed. The aim of the present work is to give a very short and elementary proof for the non-bijective version of Wigner's theorem, first for separable Hilbert spaces and then, as a consequence, for the non-separable case. 

Throughout this paper $\irH$ will denote a complex Hilbert space and $\irP_1 = \irP_1(\irH)$ will denote the set of \emph{rank-one and self-adjoint projections} on $\irH$, i.~e.~
\[ \irP_1 = \{ {\bf P}[\vec{v}] \colon \vec{v}\in\irH, \|\vec{v}\|=1 \} \] 
where ${\bf P}[\vec{v}]$ refers to the projection with precise range $\C\cdot \vec{v}$. The notations $|\vec{v}\rangle\langle \vec{v}|$ or $\vec{v}\otimes \vec{v}$ are also favourable versions for ${\bf P}[\vec{v}]$. We note that the so-called \emph{unit rays} (or \emph{pure states}) of $\irH$ and the \emph{one-dimensional subspaces} of $\irH$ can be identified with $\irP_1$ in a very natural way.

The \emph{transition probability} between two elements ${\bf P}[\vec{v}]$ and ${\bf P}[\vec{w}]$ is the quantity $\tr {\bf P}[\vec{v}]{\bf P}[\vec{w}] = |\langle \vec{v},\vec{w}\rangle|^2$. Since projections are operators on $\irH$, the usual operator-norm naturally defines a metric on $\irP_1$ i.~e.: $d({\bf P}[\vec{v}],{\bf P}[\vec{w}]) = \|{\bf P}[\vec{v}]-{\bf P}[\vec{w}]\|$. This is the so-called \emph{gap metric} (see also \cite{BJM}). The equation 
\[ \|{\bf P}[\vec{v}] - {\bf P}[\vec{w}]\| = \sqrt{1-|\langle \vec{v},\vec{w}\rangle|^2} = \sqrt{1-\tr {\bf P}[\vec{v}]{\bf P}[\vec{w}]} \] 
is very well-known. Below, we state the non-bijective Wigner theorem which will be proven here.

\smallskip

\noindent{\bf Wigner's theorem}
\emph{Let $\irH$ be a complex Hilbert space and let us consider an arbitrary mapping $f \colon \irP_1 \to \irP_1$ which preserves the transition probability, i.~e.:}
\begin{equation}\label{WP}
\tr {\bf P}[\vec{v}]{\bf P}[\vec{w}] = \tr f({\bf P}[\vec{v}])f({\bf P}[\vec{w}]) \quad (\|\vec{v}\| = \|\vec{w}\| = 1). 
\end{equation}
\emph{Then there exists a linear or antilinear isometry ${\bf W}\colon\irH\to\irH$ such that the equation}
\[ f({\bf P}[\vec{v}]) = {\bf W}{\bf P}[\vec{v}]{\bf W}^* = {\bf P}[{\bf W}\vec{v}] \]
\emph{is satisfied for every unit vector $\vec{v}\in\irH$.}

\smallskip

An easy assertion shows that (\ref{WP}) holds if and only if $f$ is an isometry with respect to the gap metric. Now, let us consider an arbitrary metric space $(X,d)$ and two subsets $D,R\subseteq X$. We say that $R$ \emph{resolves $D$} if for every two points $x_1,x_2\in D$ whenever $d(x_1,y) = d(x_2,y)$ is satisfied for all $y\in R$, we necessarily have $x_1 = x_2$ (see also \cite{BB}). For instance on the plane ($\R^2$) three arbitrary points not lying on a single line resolves the whole plane. Another example is when $D$ is a half-plane and $R$ consists of two different points of the boundary of $D$.

In Section 2 we present our proof for separable Hilbert spaces. Our strategy is the following: first, we find a set which resolves a dense subset $D\subseteq\irP_1$. Second, we show that using a linear or antilinear isometry ${\bf V}$ we can assume that our mapping acts as the identity mapping on $D$. Finally, we conclude that this mapping necessarily has to be the identity on the whole of $\irP_1$. 

In Section 3 we prove the non-separable case, and we close our letter with some concluding remarks in Section 4.


\section{Proof of the separable case}

We will denote the dimension of our separable Hilbert space $\irH$ by $N \in \N \cup \{\infty\}$. Since Wigner's theorem is trivially true for one-dimensional spaces, we always assume that $N>1$. The symbol $\N_N$ will stand for the set $\{1,2,\dots N\}$ if $N<\infty$, and for the set of natural numbers if $N = \infty$. We fix an orthonormal base: $\{\vec{e}_j\}_{j=1}^N$. Throughout the paper $v_j := \langle \vec{v},\vec{e}_j\rangle$ will denote the $j$th coordinate of a given unit vector $\vec{v}$. 

It is quite easy to see that the set
\[ D := \{ {\bf P}[\vec{v}] \colon v_j \neq 0, \; \forall \; j\in\N_N \} \subseteq \irP_1 \]
is dense in $\irP_1$ (with respect to the gap metric). Hence, if we could somehow show that $f({\bf P}[\vec{v}]) = {\bf P}[\vec{v}]$ is valid for every ${\bf P}[\vec{v}]\in D$, it would be true on the whole of $\irP_1$, since isometries are continuous. In the lemma below we give a set $R\subseteq\irP_1$ which resolves the set $D$ defined above.

\smallskip

\noindent{\bf Lemma}
Let $\irH$ be an arbitrary separable (finite or infinite-dimensional) Hilbert space. The set
\[ R = \{{\bf P}[\vec{e}_j]\}_{j=1}^N \cup \big\{{\bf P}[\tfrac{1}{\sqrt{2}}(\vec{e}_j-\vec{e}_{j+1})], {\bf P}[\tfrac{1}{\sqrt{2}}(\vec{e}_j+i \vec{e}_{j+1})]\big\}_{1\leq j<N} \]
resolves $D$.

\begin{proof} We choose two arbitrary projections ${\bf P}[\vec{v}],{\bf P}[\vec{w}]\in D$ such that $\|{\bf P}[\vec{v}] - {\bf P}[\vec{h}]\| = \|{\bf P}[\vec{w}] - {\bf P}[\vec{h}]\|$ is satisfied for every ${\bf P}[\vec{h}]\in R$. Thus, we have
\begin{equation}\label{1}
|v_j| = |w_j| \qquad (\forall\; j)
\end{equation}
\begin{equation}\label{2}
|v_j - v_{j+1}| = |w_j - w_{j+1}| \qquad (1\leq j<N)
\end{equation}
\begin{equation}\label{3}
|v_j-i v_{j+1}| = |w_j-i w_{j+1}| \qquad (1\leq j<N).
\end{equation}
Multiplying $\vec{v}$ by a complex number of modulus one, we may assume that $v_1 = w_1 (\neq 0)$ holds. Suppose that $v_k = w_k (\neq 0)$ was proven for a number $1\leq k<N$. Then (\ref{1}), (\ref{2}) and (\ref{3}) for $j=k$ gives us $v_{k+1} = w_{k+1} (\neq 0)$. This verifies that ${\bf P}[\vec{v}] = {\bf P}[\vec{w}]$ is fulfilled, and therefore $R$ resolves $D$.
\end{proof}

We note that $R$ does not resolve $\irP_1$. An easy counterexample is obtained if we put $\vec{v} = 1/\sqrt{2}\cdot (\vec{e}_1+\vec{e}_3)$ and $\vec{w} = 1/\sqrt{2}\cdot (\vec{e}_1+i \vec{e}_3)$. We also note that if somehow we could show that $f$ is the identity mapping on $R$, then necessarily $f(D)\subseteq D$ holds, and by the above lemma $f$ has to be the identity mapping on $D$, and hence on $\irP_1$. Now, we are in position to present our proof.

\begin{proof}[Proof of Wigner's theorem in the separable case]
For every $j\in\N_N$ let ${\bf P}[\vec{g}_j] = f({\bf P}[\vec{e}_j])$, then $\{\vec{g}_j\}_{j=1}^N$ has to be an orthonormal system, but it is not necessarily a base. We will denote the subspace which is generated by this system by $\irH'$. Let $\vec{v}$ be an arbitrary unit vector in $\irH$ and let $f({\bf P}[\vec{v}]) = {\bf P}[\vec{w}]$. From (\ref{WP}) we get $|v_j| = |\langle \vec{w}, g_j \rangle|$ $(j\in\N_N)$, and since $\|\vec{w}\|^2 = 1 = \|\vec{v}\|^2 = \sum_{j=1}^N |v_j|^2 = \sum_{j=1}^N |\langle \vec{w}, g_j \rangle|^2$, Parseval's identity (see \cite[Theorem I.4.13]{Co}) implies $\vec{w}\in\irH'$. We define a linear isometry ${\bf V}\colon\irH\to\irH$ by ${\bf V} \vec{e}_j = \vec{g}_j$ $(j\in\N_N)$. Obviously the mappings $f(\cdot)$ and ${\bf V}^*f(\cdot){\bf V}$ satisfy (\ref{WP}) simultaneously because ${\bf V}^*$ maps $\irH'$ isometrically onto $\irH$ and hence the equality $|\langle \vec{x}, \vec{y} \rangle| = |\langle {\bf V}^*\vec{x}, {\bf V}^*\vec{y} \rangle| \; (\vec{x},\vec{y}\in\irH', \|\vec{x}\| = \|\vec{y}\| = 1)$ is true. Moreover, the new mapping leaves the elements ${\bf P}[\vec{e}_j]$ ($j\in\N_N$) invariant. Thus from now on, without loss of generality we may assume that every ${\bf P}[\vec{e}_j]$ ($j\in\N_N$) is a fixpoint.

Let $f\big({\bf P}[\vec{v}]) = {\bf P}[\vec{w}]$, then by the latter assumption we clearly have 
\[ |w_j| = \sqrt{\tr {\bf P}[\vec{w}] {\bf P}[\vec{e}_j]} = \sqrt{\tr f({\bf P}[\vec{v}]) f({\bf P}[\vec{e}_j])} = \sqrt{\tr {\bf P}[\vec{v}] {\bf P}[\vec{e}_j]} = |v_j| \qquad (\forall\; j). \]
An easy observation from (\ref{WP}) gives us that
\[ f\big({\bf P}[1/\sqrt{2}\cdot (\vec{e}_j-\vec{e}_{j+1})]\big) = {\bf P}[1/\sqrt{2}\cdot (\vec{e}_j-\delta_{j+1} \vec{e}_{j+1})]\]
and
\[ f\big({\bf P}[1/\sqrt{2}\cdot (\vec{e}_j+i  \vec{e}_{j+1})]\big) = {\bf P}[1/\sqrt{2}\cdot (\vec{e}_j-\varepsilon_{j+1} \vec{e}_{j+1})] \]
hold for all $1\leq j < N$ with some complex numbers such that $|\delta_{j+1}| = |\varepsilon_{j+1}| = 1$. Again, using (\ref{WP}) we easily observe the following 
\[ \sqrt{2} = |\langle (\vec{e}_j-\vec{e}_{j+1}), (\vec{e}_j+i \vec{e}_{j+1})\rangle| = |\langle (\vec{e}_j-\delta_{j+1} \vec{e}_{j+1}), (\vec{e}_j-\varepsilon_{j+1} \vec{e}_{j+1})\rangle| = |1+\delta_{j+1}\overline{\varepsilon_{j+1}}|, \] 
which gives us $\delta_{j+1} = \pm i\varepsilon_{j+1}\; (j<\N)$.

If $\varepsilon_2 = i\delta_2$ is fulfilled, we define the unitary operator ${\bf U}$ with the equations ${\bf U}\vec{e}_1 = \vec{e}_1$ and ${\bf U} \vec{e}_k = \big(\prod_{j=2}^{k}\delta_{j}\big) \vec{e}_k$ ($k>1$). If $\varepsilon_2 = -i\delta_2$ is satisfied then we define the antiunitary operator ${\bf U}$ by the  previous equations. Both of $f(\cdot)$ and ${\bf U}^*f(\cdot){\bf U}$ satisfy (\ref{WP}) simultaneously. Moreover, the latter one leaves ${\bf P}[\vec{e}_j]$ ($j = 1,2,\dots$), ${\bf P}[1/\sqrt{2}\cdot (\vec{e}_j-\vec{e}_{j+1})]$ ($j<N$) and ${\bf P}[1/\sqrt{2}\cdot (\vec{e}_1+i\vec{e}_2)]$ fixed. From now on we assume that this holds for $f$. Then we also have the equalities $f({\bf P}[1/\sqrt{2}\cdot (\vec{e}_j+i\vec{e}_{j+1})]) = {\bf P}[1/\sqrt{2}\cdot (\vec{e}_j\pm i\vec{e}_{j+1})]$ for each $1<j<N$.

Now, assume that there exists an index $j>1$ for which $f({\bf P}[1/\sqrt{2}\cdot (\vec{e}_j+i\vec{e}_{j+1})]) = {\bf P}[1/\sqrt{2}\cdot (\vec{e}_j-i\vec{e}_{j+1})]$ is satisfied. We may assume that this $j$ is the first such index. First, we show that this implies 
\begin{equation}\label{contr_eq}
f({\bf P}[v_{j-1}\vec{e}_{j-1}+t\vec{e}_j+v_{j+1}\vec{e}_{j+1}]) = {\bf P}[v_{j-1}\vec{e}_{j-1}+t\vec{e}_j+\overline{v_{j+1}}\vec{e}_{j+1}]
\end{equation}
for every $t>0, v_{j-1}\neq 0,v_{j+1}\neq 0, |v_{j-1}|^2+t^2+|v_{j+1}|^2 = 1$. After that we prove that this is not possible, which verifies that $f$ is the identity mapping on $\irP_1$, and that will complete our proof. 

In order to see (\ref{contr_eq}), let $f({\bf P}[v_{j-1}\vec{e}_{j-1} + t\vec{e}_j + v_{j+1}\vec{e}_{j+1}]) = {\bf P}[w_{j-1}\vec{e}_{j-1} + s\vec{e}_j + w_{j+1}\vec{e}_{j+1}]$ where $s\geq 0$. By (\ref{WP}) we obtain $t = s$. Because the equations $|v_{j-1}| = |w_{j-1}|$, $|v_{j-1}-t| = |w_{j-1}-t|$ and $|v_{j-1}-i t| = |w_{j-1}-i t|$ are fulfilled, we easily get $w_{j-1} = v_{j-1}$. Similarly, from the equations: $|v_{j+1}| = |w_{j+1}|$, $|v_{j+1}-t| = |w_{j+1}-t|$ and $|t-i v_{j+1}| = |t+i w_{j+1}|$, we obtain $w_{j+1} = \overline{v_{j+1}}$. Therefore (\ref{contr_eq}) has to be true.

Finally, we consider two concrete vectors: $\vec{x} = \frac{-1}{2}\vec{e}_{j-1} + \frac{1}{2}\vec{e}_j + \frac{1}{\sqrt{2}}\vec{e}_{j+1}$ and $\vec{y} = \frac{i}{2}\vec{e}_{j-1} + \frac{1}{2}\vec{e}_j + \frac{i}{\sqrt{2}}\vec{e}_{j+1}$. Then from (\ref{contr_eq}) we have $\sqrt{2}/4 = |i/4 + 1/4 -i/2| = \sqrt{\tr {\bf P}[\vec{x}]{\bf P}[\vec{y}]|} = \sqrt{\tr f({\bf P}[\vec{x}])f({\bf P}[\vec{y}])} = |i/4 + 1/4 + i/2| = \sqrt{10}/4$, which is clearly a contradiction. Therefore $f$ has to be the identity mapping on $R$, and hence on the whole of $\irP_1$.
\end{proof}


\section{The proof in non-separable spaces}

Now, as a consequence we prove Wigner's theorem in the non-separable case.

\begin{proof}[Proof of Wigner's theorem in the non-separable case] Let \\ $\{\vec{e}_1,\vec{e}_2\}\cup\{\vec{e}_{\theta,j} \colon \theta\in \Theta, j\in\N, j \geq 3\}$ denote an orthonormal base in $\irH$. If we write $\vec{e}_{\theta,j}$ with $j=1$ or $2$, we mean $\vec{e}_j$. By a similar argument as in the previous section (i.~e.~using an isometry), we may assume that $f$ leaves all projections ${\bf P}[\vec{e}_{\theta,j}]$ invariant.

For a $\theta\in\Theta$ let $\irK_{\theta}$ be the subspace in $\irH$ that is generated by $\{\vec{e}_{\theta,j} \colon j\in\N\}$. It is quite obvious that if we restrict $f$ to the set $\irP_1(\irK_{\theta})\subseteq\irP_1$, we get a mapping on $\irP_1(\irK_{\theta})$ which satisfies (\ref{WP}). Using a unitary or an antiunitary operator for which every $\vec{e}_{\theta,j}$ is an eigenvector, as in the previous section, we may assume that $f$ is the identity mapping on $\irP_1(\irK_{\theta})$. It is important to note that the action of the corresponding unitary or antiunitary operator on the two-dimensional subspace $\{\alpha \vec{e}_{\xi_1}+\beta \vec{e}_{i_2}\colon \alpha,\beta\in\C\}$ is independent of the choice of $\theta$. Therefore, we may assume that for every $\theta\in\Theta$, $f$ maps on $\irP_1(\irK_\theta)$ identically.

Finally, we prove that this implies that $f$ is the identity mapping on $\irP_1$. Consider an arbitrary projection ${\bf P}[\vec{v}]\in\irP_1$ for which $v_1 \neq 0$. The set of these projections is clearly dense in $\irP_1$. Hence, if we could show that $f({\bf P}[\vec{v}]) = {\bf P}[\vec{v}]$ holds, it would show that $f$ is the identity on $\irP_1$. 

Let ${\bf P}[\vec{w}] := f({\bf P}[\vec{v}])$, $\vec{v}_\theta := \sum_{j=1}^\infty \langle \vec{v},\vec{e}_{\theta,j}\rangle \vec{e}_{\theta,j}$ and $\vec{w}_\theta := \sum_{j=1}^\infty \langle \vec{w},\vec{e}_{\theta,j}\rangle \vec{e}_{\theta,j}$. Then we infer
\[ |v_{\theta,j}| = |w_{\theta,j}| \quad (\theta\in\Theta, j\in\N) \]
and
\[ |\langle \vec{v}, \vec{v}_\theta \rangle| = |\langle \vec{w}, \vec{v}_\theta\rangle| = |\langle \vec{w}_\theta, \vec{v}_\theta\rangle| \quad (\theta\in\Theta). \]
Since 
\[ \sum_{j=1}^\infty |v_{\theta,j}|^2 = |\langle \vec{v}, \vec{v}_\theta \rangle| = |\langle \vec{w}_\theta, \vec{v}_\theta\rangle| = \Big|\sum_{j=1}^\infty |v_{\theta,j}|^2\exp(it_{\theta,j})\Big|, \]
where $w_{\theta,j} = v_{\theta,j}\exp(it_{\theta,j})$, there exist a $\lambda_\theta\in\C, |\lambda_\theta|=1$ such that $\vec{w}_\theta = \lambda_\theta \vec{v}_\theta$ holds. Since the $1$st coordinates of the vectors $\vec{v}_\theta$ are non-zero and they coincide, and the same is also true for the vectors $\vec{w}_\theta$, we get that $\vec{v} = \lambda \vec{w}$ holds with a $\lambda\in\C, |\lambda|=1$. But this implies that $f$ is indeed the identity on the above mentioned dense subset of $\irP_1$ and hence on the whole of $\irP_1$.
\end{proof}


\section{Concluding remarks}

We provided a completely new proof for Wigner's theorem. The advantages of our approach is that it is short, there are no hard calculations, it is very elementary and it works for the general case. As far as we know, such a proof which shares all of these advantages was never given before.


\section{Acknowledgements}

The author is very grateful to professor L. Moln\'ar, Dr.~G. Nagy, P.~Szokol and T.~F.~G\"orbe for some important observations and suggestions which helped to improve this paper.

The author was supported by the "Lend\"ulet" Program (LP2012-46/2012) of the Hungarian Academy of Sciences.\\ This research was also supported by the European Union and the State of Hungary, co-financed by the European Social Fund in the framework of T\'AMOP-4.2.4.A/2-11/1-2012-0001 'National Excellence Program'


\bibliographystyle{amsplain}

\end{document}